\documentclass[aps,prd,amssymb,nofootinbib,superscriptaddress,twocolumn]{revtex4}
\usepackage{graphicx}
\usepackage{amsmath}
\usepackage{amssymb}
\usepackage{amsfonts}
\usepackage{bbm}
\usepackage{amsbsy}
\usepackage{color}
\usepackage{cancel}
\usepackage{multirow}
\usepackage{soul}
\usepackage{ulem} 
\usepackage{marginnote}
\usepackage{tikz}
\usepackage{dsfont}
\usepackage{hyperref}

\begin{document}

\title{Quantum communications and quantum metrology in the spacetime of a rotating planet}

\author{Jan Kohlrus}
\affiliation{School of Mathematical Sciences,
University of Nottingham,
University Park,
Nottingham NG7 2RD,
United Kingdom}

\author{David Edward Bruschi}
\affiliation{Racah Institute of Physics and Quantum Information Science Centre, the Hebrew University of Jerusalem, Givat Ram, 91904 Jerusalem, Israel}
\affiliation{York Centre for Quantum Technologies, Department of Physics, University of York, Heslington, YO10 5DD York, United Kingdom}

\author{Jorma Louko}
\affiliation{School of Mathematical Sciences,
University of Nottingham,
University Park,
Nottingham NG7 2RD,
United Kingdom}

\author{Ivette Fuentes}
\affiliation{School of Mathematical Sciences,
University of Nottingham,
University Park,
Nottingham NG7 2RD,
United Kingdom}
\affiliation{Faculty of Physics, University of Vienna, Boltzmanngasse 5, 1090 Wien, Austria}

\begin{abstract}
We study how quantum systems that propagate in the spacetime of a rotating planet are affected by the curved background. Spacetime curvature affects wavepackets of photons propagating from Earth to a satellite, and the changes in the wavepacket encode the parameters of the spacetime. This allows us to evaluate quantitatively how quantum communications are affected by the curved spacetime background of the Earth and to achieve precise measurements of Earth's Schwarzschild radius and equatorial angular velocity. We then provide a comparison with the state of the art in parameter estimation obtained through classical means. Satellite to satellite communications and future directions are also discussed.
\end{abstract}
\maketitle

\section{Introduction}

Quantum communications is a rapidly growing field which promises several technical improvements to current classical communications. One example is the use of quantum cryptography which would make communications more secure, thanks to more robust protocols than the classical ones \cite{QKD}. More fundamental aspects can also be studied using quantum communications. For example, the interplay between quantum physics and relativity can be probed through quantum communications between moving observers and within schemes in a curved spacetime background \cite{David1}. The results of the measurements can then be compared with the predictions obtained by theories that were developed in the overlap of quantum physics and relativity, the most well known and understood of these being Quantum Field Theory (QFT) in curved spacetime \cite{Parker_toms}.

Knowing quantitatively how quantum communications are affected by the curved spacetime background would enable to compensate undesirable relativistic effects in future quantum technologies. Precise values of the necessary corrections in such quantum communication setups can only be obtained with an accurate knowledge of the spacetime parameters. We thus need to employ techniques from the field of quantum metrology, which aims at exploiting quantum resources, such as entanglement, to estimate physical parameters \cite{metrology}. Within a standard estimation protocol, an input quantum state undergoes a transformation that encodes the parameter to be estimated. The resulting state of this transformation is then compared, by means of the fidelity, to a neighbouring state which is infinitesimally close in terms of the parameter. One can define a distance between these two states that is directly related to the Quantum Fisher Information (QFI), which in turn is directly related to the maximum precision one can obtain in an estimation scheme. A final measurement provides an estimation of the value of the parameter in a single-shot run \cite{metrology}.

Typical applications of quantum metrology range from phase estimation in quantum optics to estimating the gravitational potential with Bose Einstein Condensates (BECs) \cite{gw,BECspace}. However, when estimating relativistic parameters, gravity usually appears as an external potential, or a phase modification, which does not overcome the inherent inconsistency between quantum physics and relativity \cite{vanZoestGaaloul:2010}. Recently, this gap has been bridged and quantum field theory in curved spacetime has been employed as the core framework to compute the ultimate bounds on ultra-precise measurements of relativistic parameters. In particular, it was shown that it is possible to use the shifting induced on the frequency distribution of single photons ascending the gravitational potential of a static planet to estimate with great precision the distance between a user based on Earth and one on a satellite \cite{David1,David2}. In this case, gravity isn't affecting the quantum state as the simple addition of a phase. The effects due to curved spacetime can therefore not be explained by a simple \textit{ad hoc} implementation of proper time in a classical quantum mechanics scheme. Furthermore, it was shown that these effects can have potentially high impact on specific types of quantum key distribution (QKD) protocols \cite{David1}. 
This direction has the potential of leading towards the development of new relativistic and quantum technologies aimed at testing the predictions of quantum field theory in curved spacetime in space-based experiments with satellites.

In this work we extend the analysis carried out in previous works which investigated quantum estimation techniques in scenarios where photons are exchanged between Earth and a satellite \cite{David2}. There, the Earth was assumed as static and the effects on the propagation of the photons depend only on the Schwarzschild radius of the Earth. Here we consider a rotating planet, and we model the metric outside the mass distribution by the well known Kerr metric \cite{Visser}. The transformation induced by the curvature on the traveling photon reduces to a beam-splitter, a well known linear transformation in quantum optics \cite{quantuminfo}. We can therefore restrict ourselves to Gaussian states and employ the powerful covariance matrix formalism that allows to achieve analytical insight in scenarios that involve Gaussian states and linear unitary transformations \cite{Marian&Marian,gaussian}. In particular, we seek out the effects of rotation on previously employed entanglement-swapping protocols \cite{David1,David2}. 

We find the error bound on the equatorial angular velocity of the Earth and compare it with that achieved with cutting edge technology. The rate of improvement of quantum optical technologies and the rapid increase of the control over quantum systems suggest that in the near future our scheme might provide a reliable way to outperform current technologies based on classical means.

The paper is organised as follows. In Section \ref{formalism}, we present the process of exchanging photons between Earth and a satellite, we characterise and model the system, and we give the mathematical formalism that is going to be relevant for the general relativistic calculations that will follow. In Section \ref{freqshift}, we derive the expression of the frequency shift for the photon travelling through the Kerr spacetime. Section \ref{RQM} consists of the relativistic quantum metrology calculations. It introduces the relevant perturbative quantities that are affecting the states, and derives the Quantum Fisher Information (QFI) for the system studied, and hence the estimated error bounds for the spacetime parameters. Section \ref{sts:section} introduces the satellite to satellite scheme and the related precision estimations are computed in the same fashion as in the Earth to satellite case. Finally, Section \ref{QBER} briefly discusses how the effects computed in this work can affect a simple QKD protocol, specifically comparing the magnitude of the effect with what has been found in \cite{David1}.

Throughout the whole paper we employ geometrical units $G=1=c$. Relevant constants are restored when needed for the sake of clarity. Vectors and matrices are denoted in bold characters. Vectors are written using the usual differential geometry notation \cite{Wald}, namely $\boldsymbol{X}=(X^t, \,X^r, \,X^{\theta}, \, X^{\phi}) = X^t \, \partial_t + X^r \, \partial_r + X^{\theta} \, \partial_{\theta} + X^{\phi} \, \partial_{\phi}$. Einstein's summation convention is assumed on repeated Greek indices. $A$ and $B$ indices denote evaluations at Alice's and Bob's events respectively.

\section{Introduction to the formalism \label{formalism}}

\subsection{Description of the experiment}

In this work we consider a spherical planet that rotates slowly. The Kerr metric can be used, to good approximation, to model the spacetime background around the rotating planet \cite{Visser}. Our work will be constrained to the equatorial plane $\theta=\frac{\pi}{2}$ to be able to work with simple analytical formulas. The reduced metric in Boyer-Lindquist coordinates $(t,r,\phi)$  reads \cite{Visser}:
\begin{align}\label{metric}
ds^2=&\, -\Big(1-\frac{2M}{r} \Big)dt^2+\frac{1}{\Delta}dr^2 \nonumber \\
&\,+\Big(r^2+a^2+\frac{2Ma^2}{r}\Big) d\phi^2 - \frac{4Ma}{r} dt \, d\phi, \\
\Delta=&\,1-\frac{2M}{r}+\frac{a^2}{r^2}.
\end{align}
For clarity, we will consider the rotating planet to be the Earth, with mass $M$, radius $r_A$, angular momentum $J$ and Kerr parameter (i.e., angular momentum per unit mass) $a=\frac{J}{M}$.

\begin{figure}[h!]
\includegraphics[width=8.5cm]{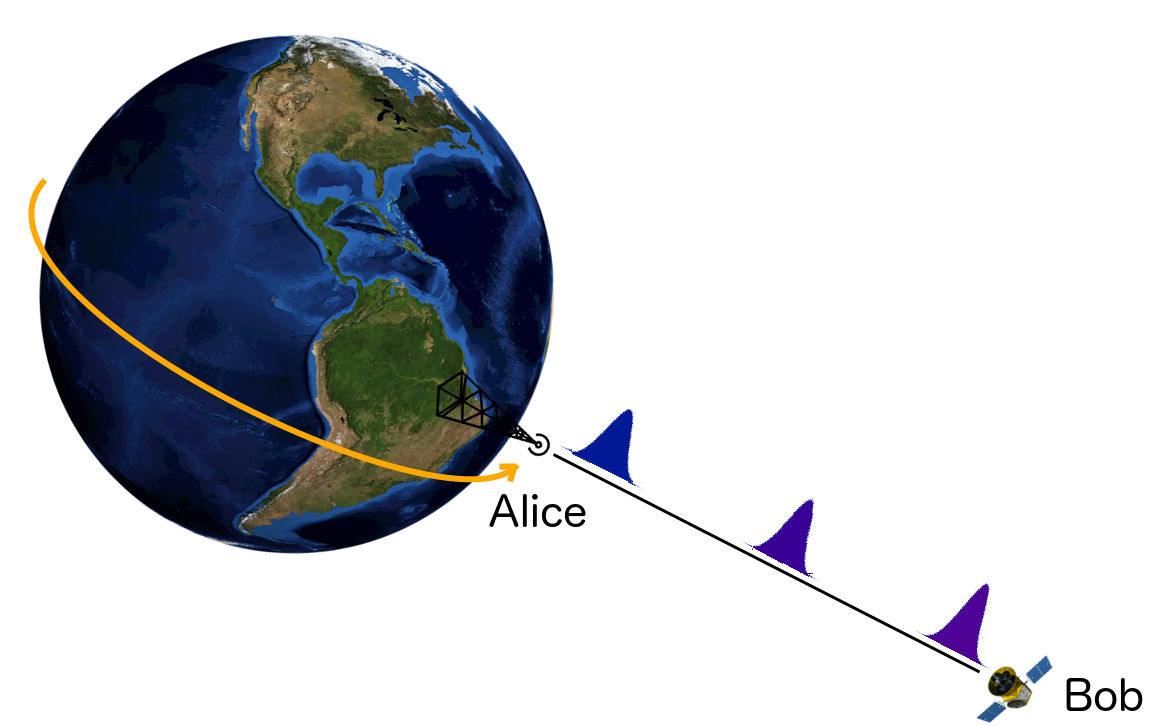}
\caption{
Alice on Earth sends a photon (localised around the straight line) to Bob in the satellite. The photon experiences the effects of the curvature of spacetime along the whole path while propagating, which can be seen in the picture by the progressive tightening and redshifting of the Gaussian wavepacket. The final effect is a nonlocal and cumulative effect due to travel along the whole path.}\label{fig:Illustration}
\end{figure}

A photon is sent radially by Alice from a laboratory on Earth's equator to Bob who is in a satellite circularly orbiting at radius $r_B$ in the equatorial plane of the Kerr spacetime. A schematic representation of the setup can be found in Fig. \ref{fig:Illustration}.

\subsection{Wave packet characterisation}

In this work we employ photons which are geometrically radial, namely with vanishing angular velocities $\dot{\phi}_{\gamma}=0=\dot{\theta}_{\gamma}$. We will see that in Kerr space-time such photons have a non trivial angular momentum constant of motion, contrarily to the Schwarzschild case. The evolution of the quantum field is thus a 1+1 dimensional problem. Such a photon can be modelled by a wave packet of frequency distribution $F_{\omega_0}(\omega)$ of monochromatic plane waves with frequency $\omega$ and peaked at $\omega_0$. The annihilation operator associated to this photon by an observer (infinitely) far from Earth is:
\begin{eqnarray}
a_{\omega_0}(t)=\int_0^{+\infty}d\omega\, e^{-i\omega t} F_{\omega_0}(\omega)\,a_{\omega}.\label{wave:packet:infinity}
\end{eqnarray}
The canonical bosonic commutation relations $[a_{\omega_0}(t),a^{\dagger}_{\omega_0}(t)]=1$ for the bosonic operator \eqref{wave:packet:infinity} at any instant of time follow directly from those for the sharp frequency operators $[a_{\omega},a^{\dagger}_{\omega^{\prime}}]=\delta(\omega-\omega^{\prime})$ and from the normalisation of the frequency distribution function $F_{\omega_0}(\omega)$ i.e., $\int_{0}^{\infty} d\omega |F_{\omega_0}(\omega)|^2=1$.

It is possible to rewrite the annihilation operator \eqref{wave:packet:infinity} of the same photon as described by Alice or Bob. We then follow notation in \cite{David1} and reproduce the equation below:
\begin{eqnarray}
a_{\Omega_{K,0}}(\tau_K)=\int_0^{+\infty}d\Omega_K\, e^{-i\Omega_K \tau_K} F^{(K)}_{\Omega_{K,0}}(\Omega_K)\,a_{\Omega_K},\label{wave:packet}
\end{eqnarray}
where the index $K=A,B$ refers to observer Alice or Bob respectively. The quantity $\Omega_K$ is the frequency of the photon as measured locally by the observer $K$ with proper time $\tau_K$. We have introduced the peak frequency $\Omega_{K,0}$ of the frequency distribution $F^{(K)}_{\Omega_{K,0}}$, and the bosonic canonical commutation relations  for each observer read $[a_{\Omega_K},a^{\dagger}_{\Omega_K}]=1$. 

Alice now prepares and sends a wave packet $F^{(A)}_{\Omega_{A,0}}$ at altitude $r_A$ which is received by Bob on the satellite as a wave packet $F^{(B)}_{\Omega_{B,0}}$ at altitude $r_B$. Throughout its journey, the wave packet has changed due to the spacetime being curved. The relation between the two frequency distributions has been already found in \cite{David1}. We define the frequency shift as $\Omega_B=f \Omega_A$, where $f$ is the \textit{total} shifting function that will be made explicit later (notice that we are not using the same definition for $f$ as in \cite{David1,David2}). Then one finds:
\begin{eqnarray}
F^{(B)}_{\Omega_{B,0}}(\Omega_B)=\sqrt{\frac{1}{f}}\,F^{(A)}_{\Omega_{A,0}}\left(\frac{1}{f} \, \Omega_B\right).\label{wave:packet:relation}
\end{eqnarray}
Bob knows that the photon Alice has sent was characterised by $F^{(A)}_{\Omega_{A,0}}$. One way to quantify the change in the state of the photon is to use the fidelity between the initial state prepared with wave packet $F^{(A)}_{\Omega_{A,0}}$ and the final state received with wave packet $F^{(B)}_{\Omega_{B,0}}$. The fidelity $\mathcal{F}=| \Theta |^{2}$ for a single photon in a pure state is simply defined through the overlap function between the two frequency distributions:
\begin{eqnarray}
\Theta=\int_0^{+\infty}d\Omega_B\,F^{(B)\star}_{\Omega_{B,0}}(\Omega_B)F^{(A)}_{\Omega_{A,0}}(\Omega_B).\label{single:photon:overlap}
\end{eqnarray}
The fidelity would tend to zero for photons traversing regions of the spacetime with sufficiently strong curvature, while it would reach unity in flat spacetime.

A convenient choice of wave packet is a normalised Gaussian wave packet of width $\sigma_K$ and with a frequency distribution of the form:
\begin{eqnarray}
F_{\Omega_{K,0}}^{(K)}(\Omega_{K})=\frac{1}{\sqrt[4]{2\pi\sigma_K^2}}e^{-\frac{(\Omega_{K}-\Omega_{K,0})^2}{4\sigma_K^2}}\label{Gaussian:wave:packet}.
\end{eqnarray}
We therefore employ \eqref{single:photon:overlap} and \eqref{Gaussian:wave:packet} (also see \cite{David1}) to find:
\begin{eqnarray}
\Theta=\sqrt{\frac{2(1+\delta)}{1+(1+\delta)^2}}e^{-\frac{\delta^2\Omega_{B,0}^2}{4(1+(1+\delta)^2)\sigma_B^2}}\label{overlap},
\end{eqnarray}
with the amount of shifting being quantified by the new parameter $\delta$ defined by\footnote{In this updated version, we adopt the same convention for the parameter $\delta$ as in \cite{David2} for consistency with this previous work.}:
\begin{eqnarray}
\delta=f-1. \label{delta}
\end{eqnarray}
In the following, we will derive the explicit formula for $f$ in terms of the spacetime parameters.

\section{Frequency shift in Kerr spacetime \label{freqshift}}

\subsection{Preliminaries}

The general frequency shift formula for a photon emitted from Alice on Earth and received by Bob in the satellite reads \cite{Wald,Schr}:
\begin{align}\label{frequency:formula}
f=\frac{\Omega_{B}}{\Omega_{A}} = \frac{\Big[\boldsymbol{k}.\dot{\boldsymbol{X}}_{B}\Big]_{|_{\boldsymbol{X} = \boldsymbol{X}_{B}}}} {\Big[\boldsymbol{k}.\dot{\boldsymbol{X}}_{A}\Big]_{|_{\boldsymbol{X} = \boldsymbol{X}_A}}},
\end{align}
where $\dot{\boldsymbol{X}}_A$ and $\dot{\boldsymbol{X}}_B$ are the four-velocities of Alice and Bob respectively, while $\boldsymbol{k}$ is the tangent vector to the affinely parametrised null geodesic that the photon follows. For simplicity of the computations, we restrain our study to a satellite that follows a circular orbit, i.e. we have $\dot{r}_{B} = 0 = \dot{r}_{A} $, where the dot stands for derivative with respect to proper time. 
Bob's satellite has its motion constrained to the equatorial plane of the Kerr spacetime, thus $\dot{\theta}_B=0$ and Alice has neither a $\theta-$motion. Also our photon is geometrically radial, hence:
\begin{align}
k^{\mu}(\dot{X}_{K})_{\mu} =&\,  k^{t}(\dot{t}_{K} \, g_{tt} +\dot{\phi}_{K} \, g_{t\phi}),
\end{align}
where again $K=A,B$.

The velocity of our observers are \cite{Felice_bini, Chandra1983}:
\begin{align}\label{Alice}
\dot{\boldsymbol{X}}_A=&\, \gamma_A \big(\, \partial_t + \omega_A \, \partial_{\phi} \big), \\ \label{Bob}
\dot{\boldsymbol{X}}_B=&\, \gamma_B \big( (1+ \epsilon \, a \, \omega_B) \, \partial_t + \epsilon \, \omega_B \, \partial_{\phi} \big),
\end{align}
where $\epsilon = +1$ for direct orbits (i.e., when the satellite co-rotates with the Earth), and $\epsilon = -1$ for retrogade ones (i.e., the opposite way). The parameter $\omega_A=d \phi_A / d t_A$ denotes Earth's angular velocity at the equator, while $\omega_B = \sqrt{M/r_B^3}$ is Bob's orbit frequency. The normalisation factors $\gamma_A$ and $\gamma_B$ are given by:
\begin{align}\label{gamma}
\gamma_A=&\, \Big( 1-\omega_A^2 \big(r_A^2+a^2 \big) -\frac{2M}{r_A} \big(1-a \, \omega_A \big)^2 \Big)^{-\frac{1}{2}}, \\
\gamma_B =&\, \Big(1-\frac{3M}{r_B}+2 \, \epsilon \, a \, \omega_B \Big)^{-\frac{1}{2}}.
\end{align}
The tangent vector to the photon's worldline reads:
\begin{align}\label{photon}
\boldsymbol{k}=&\, E_{\gamma} \left( \frac{1}{1-\frac{2M}{r}} \, \partial_t + \sqrt{\kappa} \, \partial_{r} \right), \\ 
\kappa =&\, 1+\frac{a^2}{r^2} \Big(1+\frac{2M}{r}\Big) + \frac{4}{1-\frac{2M}{r}}\frac{M^2 \, a^2}{r^4},
\end{align}
where it has been used that for such a geometrically radial photon we have:
\begin{align}\label{conserved quantities}
L_{\gamma}= -a \frac{E_\gamma}{1-\frac{2M}{r}} \frac{2M}{r}.
\end{align}
The constants of motion $E_{\gamma}$ and $L_{\gamma}$ are respectively the energy and longitudinal angular momentum of the photon as measured by an inertial observer at space infinity. These quantities are conserved along geodesics thanks to the presence of the two Killing fields $\partial_t$ and $\partial_{\phi}$.
After evaluation at $\boldsymbol{X}_B$, the explicit form of the numerator in \eqref{frequency:formula} thus reads:
\begin{align}\label{frequency:formula:numerator}
[k^{\mu}(\dot{X}_B)_{\mu}]_{|_{\boldsymbol{X}_B}} = -E_{\gamma} \, \gamma_B \left( 1+ \epsilon \frac{a \, \omega_B}{1-\frac{2M}{r_B}} \right).
\end{align}
The denominator of \eqref{frequency:formula}, after evaluation at $\boldsymbol{X}_A$, reads:
\begin{align}\label{frequency:formula:denominator}
[k^{\mu} (\dot{X}_{A})_{\mu}]_{|_{\boldsymbol{X}_A}} = -E_{\gamma} \, \gamma_A \left( 1+\frac{2M}{r_A}\frac{a \, \omega_A}{1 - \frac{2M}{r_A}} \right).
\end{align}
The $A$ and $B$ subscripts on the quantities $\Delta$ and $\kappa$ denote evaluation at $r_A$ and $r_B$ respectively. We now have all the ingredients to compute explicitly the frequency shift of the photon \eqref{frequency:formula}.

\subsection{Frequency shift formula}

Plugging \eqref{frequency:formula:numerator} and \eqref{frequency:formula:denominator} in \eqref{frequency:formula}, we obtain the explicit expression of the frequency shift for the photon exchanged between Alice on Earth and Bob in the circularly orbiting satellite. We find:
\begin{align}\label{freqshiftrad}
f =&\, \frac{1+ \epsilon  \frac{ a \, \omega_B}{1-\frac{2M}{r_B}}}{1+\frac{2M}{r_A} \frac{ a \, \omega_A}{1-\frac{2M}{r_A}}} \,  \sqrt{\frac{ 1-\frac{2M}{r_A}(1-a \, \omega_A)^2 - (a^2 + r_A^2) \omega_A}{1-\frac{3M}{r_B}+ 2 \, \epsilon \, a \, \omega_B}}.
\end{align}
In the Schwarzschild limit $(a, \omega_A) \to (0,0)$, the frequency shift simplifies to:
\begin{align}\label{schwrad}
f_S = \sqrt{\frac{1-\frac{2M}{r_A}}{1-\frac{3M}{r_B}}}.
\end{align}
Therefore, equation \eqref{freqshiftrad} reduces to the known result for a radial photon in a static planet spacetime that has been used in \cite{David1,David2}. As expected by the symmetry of the problem in the case of a radial photon propagating in Schwarzschild spacetime, the result does not depend on the direction of rotation of the satellite, namely on $\epsilon$. One can also notice that, in this limit, photons received on satellites orbiting at radius $r_B=\frac{3}{2}r_A$ will not experience any frequency shift. In the Schwarzschild picture, this is the altitude at which the gravitational effect of the Earth and the special relativistic effect due to the motion of the satellite compensate each other, and Bob's clock rate becomes equal to the clock rate of Alice. Indeed, the satellite's motion around the Earth slows down Bob's proper time, but the higher altitude of Bob introduces a lower redshift which therefore has also a lower effect on Bob's clock rate, as compared to Alice. Special relativistic effects thus dominate the frequency shift of photons received at altitudes below $r_B=\frac{3}{2}r_A$, where photons will actually be received blue-shifted, while the photons will be received red-shifted at $r_B>\frac{3}{2}r_A$ where the gravitational frequency shift dominates. A last relevant check is to verify the absence of frequency shift in flat spacetime. Unsurprisingly, we get from the relevant limit of \eqref{freqshiftrad} that in Minkowski spacetime $f_{M}=1$.

\section{Quantum estimation of rotation parameters of the Earth \label{RQM}}

In this section, we apply quantum estimation techniques to find the ultimate bounds on the precision of measurements of parameters of the Earth.

\subsection{Summary of spacetime parameters}

In our result \eqref{freqshiftrad} for the frequency shift of a radial photon traveling from Earth to space there are five dimensionless perturbative parameters of interest, for which we give numerical values in the table \ref{paramtable}.

\begin{table}[!htbp]
\begin{tabular}{| c | c | c | c |}
\hline
Quantity (N. Units) & Quantity (S.I.)  & Value & Orbit \\ \hline
$M/r_A$ & $GM/(r_Ac^{2})$ & $6.95 \times 10^{-10}$ & / \\ \hline
\multirow{2}{*}{$M/r_B$} & \multirow{2}{*}{$GM/(r_Bc^{2})$} & $1.05 \times 10^{-10}$ & GEO  \\ 
 & & $5.29 \times 10^{-10}$ & LEO  \\ \hline
$a/r_A$ & $a/r_A$ & $ 5.11 \times 10^{-7}$ & / \\ \hline
 \multirow{2}{*}{$a/r_B$} & \multirow{2}{*}{$a/r_B$} & $7.73 \times 10^{-8}$ & GEO \\
 & & $3.89 \times 10^{-7}$ & LEO  \\ \hline
 $r_A\omega_A$ & $r_A\omega_A/c$ & $1.55 \times 10^{-6}$ & / \\ \hline
\end{tabular}
\caption{Dimensionless perturbative parameters in the frequency shift formula.}
\label{paramtable}
\end{table}

We have used the following values: $a = 3.26$m, $\omega_A=7.29 \times 10^{-5}$rad/s, $r_A=6378$km, $M=5.97 \times 10^{24}$kg. Furthermore we consider two orbits for satellites, low Earth orbits $r_B(\mathrm{LEO}) = r_A + 2000$km and geostationary ones $r_B(\mathrm{GEO}) = r_A + 35784$km.

\subsection{Quantification of the frequency shift}

The amount of change in the photon's frequency distribution is quantified by our general parameter $\delta$: 
\begin{align}
\delta = \frac{\Omega_{B}}{\Omega_{A}}-1,
\end{align}
where $\Omega_{B}/\Omega_{A}$ has an explicit expression in \ref{freqshiftrad}.
We proceed by expanding perturbatively \eqref{freqshiftrad} in the parameters from table \ref{paramtable}. We obtain a $\delta$ parameter of the following form:
\begin{align}\label{shiftquant}
\delta =&\, \delta_S + \delta_{rot} + \delta_{c},
\end{align}
where $\delta_S$ is a first order Schwarzschild term, $\delta_{rot}$ is the lowest order rotation term, and we gather all higher order corrections in $\delta_c$. We give explicit values of the first two:
\begin{align}
\delta_{S} =&\, \frac{1}{4}\frac{r_{S}}{r_A}\frac{1-2\frac{L}{r_A}}{1+\frac{L}{r_A}}, \label{delta:S} \\
\delta_{rot} =&\, -\frac{(r_A \, \omega_A)^2}{2} \sim -10^{-12},
\end{align}
where we have introduced the Schwarzschild radius of the Earth $r_S=2M$ and the distance between Alice and Bob $L=r_B-r_A$. Notice that $\delta_S$ in \eqref{delta:S} is different to the $\delta$ displayed in the Schwarzschild scenario papers \cite{David1,David2}. It comes from the fact that we are expanding the \textit{total} frequency shift \eqref{freqshiftrad} taking into account both special and general relativistic effects, while in \cite{David1,David2} $\delta$ has been obtained by expanding only the gravitational frequency shift. With the values used in table \ref{paramtable}, we have $\delta_S \sim 10^{-10}$ for LEO orbits and $\delta_S \sim -10^{-10}$ for GEO orbits. Lowest order terms from $\delta_c$ are of order $10^{-21}$.

If one assumes that the positions of Alice and Bob are known with sufficient (i.e. infinite) precision, one can employ \eqref{shiftquant} to express the precision $\Delta \delta$ on measurements on $\delta$ in terms of the precision $\Delta r_S$ on the Schwarzschild radius:
\begin{align}\label{error:relation:delta:schwarzschild:radius}
\Delta \delta = \delta_S \frac{\Delta r_{S}}{r_S}.
\end{align}
We have used that for most orbits $|\delta_S| \gg |\delta_c|$ to neglect terms coming from $\delta_c$. Yet, as noticed in \eqref{schwrad}, there is no frequency shift in Schwarzschild spacetime for orbits $L=r_A/2$, and consequently $\delta_{S}$ vanishes for these orbits. Hence, for such orbits $L\sim r_A/2$ we need to take higher order corrections from $\delta_c$ into account, and \eqref{error:relation:delta:schwarzschild:radius} will have a more involved expression.

We are also interested in the precision one can achieve for the measurement of Earth's equatorial angular velocity. The relation between $\Delta \delta$ and $\Delta \omega_A$ is simply:
\begin{align}\label{error:relation:delta:angular:velocity}
\Delta \delta =&\, 2\delta_{rot} \frac{\Delta \omega_A}{\omega_A}.
\end{align}
We will now proceed to employ the quantum estimation techniques necessary to find the ultimate bounds on the measurement errors we have explicitly found above.
In order to achieve this goal, we need to compute the Quantum Fisher Information $\mathrm{H(\delta)}$ which will allow us to employ the Cram\'er-Rao theorem.

\subsection{Quantum Fisher Information (QFI) and single parameter estimation}

The most important quantity in quantum metrology is the quantum Fisher information $\mathrm{H}$, which allows to directly compute the bounds on measurements of interest through the well known Cram\'er-Rao theorem \cite{Cramer}. In particular it was shown in  \cite{Cramer} that, if one wishes to estimate the parameter $\delta$ encoded in the final state $\rho(\delta)$ of a system after a transformation $U(\delta)$, one can employ the fidelity $\mathcal{F}(\rho(\delta),\rho(\delta+d\delta))$ between the states $\rho(\delta)$ and $\rho(\delta+d\delta)$ and obtain the QFI as
\begin{align}
\mathrm{H}=\lim_{d\delta \to 0} 8 \, \frac{1-\sqrt{\mathcal{F}(\rho(\delta),\rho(\delta+d\delta))}}{d\delta^2}.
\end{align}
One can then compute the ultimate bound on the error $\Delta\delta$ through the Cram\'er-Rao inequality as 
\begin{align}
| \Delta\delta | \geq\frac{1}{\sqrt{N \, \mathrm{H}}},
\end{align}
where $N$ is the number of probes in the experiment.

Following \cite{David2}, we employ an initial two mode squeezed state and compute the fidelity $\mathcal{F}$ in order to obtain the QFI.
For the regime we are interested in, $\delta\ll \frac{\Omega^2}{8 \sigma^2}\delta^2 \ll1$, it reads:
\begin{align}
\mathcal{F} = 1 -  \frac{\Omega_{1}^{2}+ \Omega_{2}^{2}}{4\sigma^{2}} \sinh^{2}(s) \, d\delta^{2},
\end{align}
where $s$ denotes the squeezing parameter, $\sigma$ the spread of the frequency distribution of the photon, and $\Omega_{i}$ denote the peak frequencies of the distribution of each mode, i.e., $i=1,2$. From this we compute the QFI as:
\begin{align}
\mathrm{H} = \frac{\Omega_{1}^{2}+ \Omega_{2}^{2}}{\sigma^{2}} \sinh^{2}(s).
\end{align}
Finally, we find our desired result:
\begin{align}\label{final:result}
|\Delta\delta| \geq \frac{\sigma}{\sqrt{N(\Omega_{1}^{2}+ \Omega_{2}^{2})}\sinh(s)}.
\end{align}
In the following we specialise equation \eqref{final:result} to different estimations, such as estimation of the Schwarzschild radius or the equatorial angular velocity of the Earth.

\subsection{Optimal bounds for the error on spacetime parameters}

In this section we will focus on applying the previous techniques to estimate the ultimate error bounds on the Schwarzschild radius $r_S$ and on the equatorial angular velocity $\omega_A$ of the Earth. We assume absence of losses and use typical values for the parameters of the setup such as the bandwith $\sigma = 10^{6}\mathrm{Hz}$, the peak frequencies $\Omega_1=\Omega_2=\Omega=7 \times 10^{14}\mathrm{Hz}$ and the allowed number of measurements $N=10^{10}$. In practice, these numbers imply a measurement time of $N \sigma^{-1} \sim 3\mathrm{hours}$. Furthermore, we present results for squeezing $s=2$, which is achievable with state-of-the-art technology \cite{David2,squeezing}. 
The optimal bound for the error on the measurement of Earth's Schwarzschild radius is given by:
\begin{align}\label{schwmes}
\frac{|\Delta r_{S}|}{r_{S}} \geq \frac{1}{\sqrt{2N}\sinh(s)} \frac{\sigma}{\Omega} \big|\delta_S\big|^{-1}.
\end{align}
The rotation terms being negligible, the result is essentially the same bound as in \cite{David2} for measurements of the Schwarzschild radius, namely $|\Delta r_S/r_S| \sim 10^{-5}$ for LEO orbits and $|\Delta r_S/r_S| \sim 10^{-6}$ for GEO orbits. Yet, these values now take into account special relativistic effects due to Alice's and Bob's motion.

For orbits at altitude around $L\sim\frac{r_A}{2}$ however, the Schwarzschild term $\delta_S$ in \eqref{schwmes} vanishes, we then need to add the lowest order terms from $\delta_c$. These are several orders of magnitude smaller than $\delta_S$, therefore satellites orbiting at these altitudes are not recommended for the experiments proposed here since the precision they would provide for the measurement of the Schwarzschild radius is significantly lower. This result is new compared to the study carried in \cite{David2}, it comes from taking into account special relativistic effects due to our observers' motions.

We shift our attention to estimating the bound for the equatorial angular velocity of the Earth. We get:
\begin{align}
\frac{|\Delta \omega_A|}{\omega_A} \geq \frac{1}{2\sqrt{2N}\sinh(s)} \frac{\sigma}{\Omega} \big|\delta_{rot}\big|^{-1},
\end{align}
which gives bounds of order $|\Delta \omega_A|/\omega_A\sim 10^{-3}$. We are five orders of magnitude below the IERS Numerical Standards that give a relative uncertainty of order $10^{-8}$ \cite{IAG1999}, as well as the per billion precision of a recent direct measurement involving large ring laser gyroscopes \cite{newexp} and of old interferometer experiments \cite{oldexp}. 
However, given the current rate of improvement in quantum technologies, it is reasonable to assume that, in the near future, we will be able to employ higher squeezing values and photons of higher energy. Finally, larger number of measurement probes would also contribute to enabling us to exceed the state-of-the-art precision.

\section{Satellite to satellite communication\label{sts:section}}

Another possible experimental setup would see two parties, Bob and Charlie, both following geodesic circular orbits in the equatorial plane of the Earth, located at altitudes $r_B$ and $r_C$ respectively, with $r_B>r_C$.  The advantage of this setup is that the channel (i.e., the free space between the two parties) is free from the noise introduced, for example, by the presence of the atmosphere in the case of Alice sending a photon from Earth to Bob's satellite \cite{Bonato:Tomaello:2009,Vallone:Bacco:2015}.
Using \eqref{frequency:formula} with Charlie instead of Alice and \eqref{frequency:formula:numerator} for Bob and Charlie, the general frequency shift formula for a photon emitted from Charlie's device on a satellite and received later by Bob on a higher satellite reads:
\begin{align}\label{freqshift:sat2sat}
\frac{\Omega_{B}}{\Omega_{C}} =&\, \frac{1+ \epsilon  \frac{ a \, \omega_B}{1-\frac{2M}{r_B}}}{1+ \eta  \frac{ a \, \omega_C}{1-\frac{2M}{r_C}}} \, \sqrt{\frac{1-\frac{3M}{r_C}+ 2 \, \eta \, a \, \omega_C}{1-\frac{3M}{r_B}+ 2 \, \epsilon \, a \, \omega_B}},
\end{align}
where all the quantities with a $C$ subscript are the same as Bob's but substituting $r_B$ with $r_C$ and $\epsilon$ with $\eta$. Similarly to Bob's $\epsilon$, $\eta=\pm1$ depending on which way Charlie's satellite revolves around the Earth.
In this expression there are four perturbative parameters of interest: the Schwarzschild parameters $M/r_B$, $M/r_C$, and the Kerr parameters $a/r_B$, $a/r_C$. In order to obtain the shift quantity $\delta_{s} = (\Omega_{B}/\Omega_{C})-1$ that quantifies the shift in the frequency distribution of the photon, we need to expand perturbatively the square root of \eqref{freqshift:sat2sat} with respect to these four parameters. Doing so, we find an expression of the following form:
\begin{align}\label{delta:sats}
\delta_{s} = \delta_{s, S} + \delta_{s, rot} + \delta_{s,c},
\end{align}
with:
\begin{align}
\delta_{s, S} =&\, -\frac{3}{4}\frac{L \, r_S}{r_C^2} \frac{1}{1+\frac{L}{r_C}} \sim -10^{-10}, \label{delta:sats:S} \\
\delta_{s, rot} =&\, \frac{1}{4} \frac{r_S \, a^2}{r_C^3}\bigg(\Big(1+\frac{L}{r_C}\Big)^{-3}-1\bigg)\sim -10^{-23},\label{delta:sats:rot}
\end{align}
where now $L=r_B-r_C>0$, $\delta_{s,c}$ are higher order contributions that are negligible, and we give numerical values of $\delta_{s, S}$ and $\delta_{s, rot}$ for Charlie following a LEO and Bob a GEO. Notice that in this scheme where both observers are geodesic, contrary to the Earth to satellite setup, there are no orbits for which the Schwarzschild term $\delta_{s,S}$ vanishes. We can now express the error $\Delta \delta_{s}$ on our shift parameter $\delta_{s}$ in terms of the error on the spacetime parameters. We find:
\begin{align}
\Delta\delta_{s} =&\, (\delta_{s, S} + \delta_{s, rot} ) \frac{\Delta r_S}{r_S} \approx  \delta_{s, S} \frac{\Delta r_S}{r_S}, \\
\Delta\delta_{s} =&\, 2 \delta_{s, rot} \frac{\Delta a}{a}  = 2 \delta_{s, rot} \frac{\Delta \omega_A}{\omega_A},
\end{align}
where we have used for the last equality that $a = 2 I \omega_A / r_S$, where $I$ is Earth's moment of inertia. For a photon sent from Charlie on a low Earth orbit ($r_C \sim 8000$km) to Bob on a geostationary one ($r_B \sim 42000$km), we find the order of magnitude for the precision on the Schwarzschild radius to be $|\Delta r_S|/r_S \sim 10^{-6}$ and on the rotation parameters $|\Delta a| / a \sim |\Delta \omega_A| / \omega_A \gg 1$. Therefore, in this satellite to satellite scheme, the rotation parameters measurements are losing several orders of magnitude of precision compared to the Earth to satellite setup. This is understandable by looking at the nature of the observer. On Earth, Alice is strongly dragged by Earth's rotation while the satellites experience only a slight dragging due to the weak rotation of the metric. It is then not surprising that the satellite to satellite setup, which is made of two geodesic orbiting observers, is less sensitive to the rotation parameters of the Kerr spacetime. However, the value for the precision on the measurement of the Schwarzschild radius is similar to the Earth to satellite scheme, making both setups equally good in theory. Yet, one has to keep in mind that a satellite to satellite scheme will provide channels free from any atmospheric noise and should therefore eventually yield more precise measurements.

\section{Quantum Bit Error Rate (QBER) in a simple QKD protocol \label{QBER}}

In order to complete our analysis of the possible means of detecting these effects, we can compute the QBER for a simple QKD protocol, following closely what has been done in \cite{David1}. Alice and Bob have two memories each: $A1, A2$ and $B1, B2$ respectively. The optical modes contained in the memories of one user (e.g. Alice's) are propagated to the other user (Bob, in this case). The optical modes from memories $A1$ and $B1$ are then entangled at the receiver's lab and similarly for $A2$ and $B2$. Alice then beam splits $A1$ and $A2$ and each output branch of the beamsplitter is measured by a detector. Bob performs the same operation with $B1$ and $B2$. If each user has one detector clicking, the protocol has been working successfully. The probability for Alice and Bob to share the same bit, i.e., the probability for memories $A1, B1$ and $A2, B2$ to have the same state is $p=1-q/2$, where $q \ll 1$ will in our case be related to our $\delta$ parameter and the wave packet distributions. The QBER is the rate of bits that were not shared between Alice and Bob, i.e. $\mathrm{QBER}=\bar{p}=1-p=q/2$. We employ the same protocol between Alice and Bob and adapt it to our new results, which take into account Earth's rotation and special relativistic effects. From \cite{David1} we have:
\begin{align}
\mathrm{QBER} \sim \frac{\delta^2}{8} \frac{\Omega^2}{\sigma^2},
\end{align}
in the regime $\delta\ll \frac{\Omega^2}{8 \sigma^2}\delta^2 \ll1$. In the Earth to satellite setup, the contribution of the rotation to $\delta$ in \eqref{shiftquant} is negligible for most orbits. We obtain a QBER of order $10^{-4}$ for communications to LEO orbits and $10^{-2}$ to GEO orbits. For orbits at radii $r_B \sim \frac{3}{2} r_A$ however, the rotation term becomes dominant and the QBER shrinks to $\sim 10^{-8}$. Hence, these orbits are recommended to reduce the QBER in Earth to satellite quantum communications.

In the satellite to satellite case, the Schwarzschild part of the shift is always dominant. The value of the shift between a LEO and a GEO satellite is similar to the GEO orbits case in the Earth to satellite scheme, hence the value for the QBER for quantum communications between a LEO and a GEO satellite is of order $10^{-2}$ too. However, taking into account atmospheric effects in the ground to satellite case would make the satellite to satellite scheme more accurate.

\section{Conclusion}

In this paper we have derived an expression for the general relativistic frequency shift of a photon travelling through Earth's rotating surrounding spacetime. We have specialised to photons travelling with vanishing angular velocities from an equatorial laboratory on Earth towards a satellite revolving in the equatorial plane of the Kerr spacetime. This study provides analytical insight and successfully extends previous results obtained for Schwarzschild spacetime \cite{David1,David2}. We have found that including the rotation of the Earth does not change previous estimates obtained for the Schwarzschild radius in a quantum metrology scheme. However, we were able to estimate the precision for the quantum measurement of the equatorial angular velocity of the Earth. We find that the error bound predicted for the equatorial angular velocity of the Earth can exceed the precision obtained with the state of the art when high values of squeezing and a large number of probe systems (or measurements) are employed. Suitably chosen signals, such as frequency comb, instead of gaussian-shaped frequency distributions, could also improve precision \cite{losses,opt_phot}. Taking into account special relativistic effects, we have also found a specific class of circular orbits where the frequency of the received photons remains almost unchanged. For quantum metrology purposes these orbits have to be avoided since the quantum state of the photons is less perturbed, yet they are very useful for minimal curved spacetime disturbance channels for quantum communication. To complete our analysis, we have added a study of the error bounds for the same parameters when communication occurs between two satellites, which has relevance for practical implementations of many quantum information schemes, such as proposed implementations of QKD through satellite nodes \cite{quantum_sats}. We conclude that recent advances in quantum technologies, which include the ability to create larger values of squeezing, show the promising opportunities of improving the state of the art for measurements of physical parameters of the Earth. 

\section*{Acknowledgments}
J. Louko was supported in part by STFC (Theory Consolidated Grant ST/J000388/1). D. E. Bruschi was partially supported by the I-CORE Program of the Planning and Budgeting Committee and the Israel Science Foundation (grant No. 1937/12), as well as by the Israel Science Foundation personal grant No. 24/12. D. E. Bruschi would also like to thank the University of Vienna for hospitality.

\bibliographystyle{unsrt}
\bibliography{Kerrpaper}

\begin{thebibliography}{10}

\bibitem{QKD}
IEEE International conference on computers, systems and signal processing.
\newblock {\em Quantum Cryptography: Public key distribution and coin tossing},
  1984.

\bibitem{David1}
David~Edward Bruschi, Ivette Fuentes, Thomas Jennewein, and Mohsen Razavi.
\newblock Spacetime effects on satellite-based quantum communications.
\newblock {\em Phys. Rev. D}, 90:045041, 2014.

\bibitem{Parker_toms}
Leonard Parker and David Toms.
\newblock {\em Quantum field theory in curved spacetime}.
\newblock Cambridge University Press, Cambridge, 2009.

\bibitem{metrology}
Vittorio Giovanetti, Seth Lloyd, and Lorenzo Maccone.
\newblock Advances in quantum metrology.
\newblock {\em Nature Photon.}, 5:222, 2011.

\bibitem{gw}
Carlos Sab\'in, David~Edward Bruschi, Mehdi Ahmadi, and Ivette Fuentes.
\newblock Phonon creation by gravitational waves.
\newblock {\em New J. Phys.}, 16:085003, 2014.

\bibitem{BECspace}
David~Edward Bruschi, Carlos Sab\'in, Angela White, Valentina Baccetti,
  Daniel~K.L. Oi, and Ivette Fuentes.
\newblock Testing the effects of gravity and motion on quantum entanglement in
  space-based experiments.
\newblock {\em New J. Phys.}, 16:053041, 2014.

\bibitem{vanZoestGaaloul:2010}
T.~van Zoest~\textit{et al.}
\newblock Bose-{E}instein condensation in microgravity.
\newblock {\em Science}, 328(5985):1540--1543, 2010.

\bibitem{David2}
David~Edward Bruschi, Animesh Datta, Rupert Ursin, Timothy~C. Ralph, and Ivette
  Fuentes.
\newblock Quantum estimation of the {S}chwarzschild spacetime parameters of the
  {E}arth.
\newblock {\em Phys. Rev. D}, 90:124001, 2014.

\bibitem{Visser}
Matt Visser.
\newblock The {K}err spacetime: A brief introduction.
\newblock arXiv:0706.0622, 2007.

\bibitem{quantuminfo}
Michael~A. Nielsen and Isaac~L. Chuang.
\newblock {\em Quantum computation and quantum information}.
\newblock Cambridge University Press, Cambridge, 2000.

\bibitem{Marian&Marian}
Paulina Marian and Tudor~A. Marian.
\newblock Uhlmann fidelity between two-mode {G}aussian states.
\newblock {\em Phys. Rev. A}, 86:022340, 2012.

\bibitem{gaussian}
Gerardo Adesso, Sammy Ragy, and Antony~R. Lee.
\newblock Continuous variable quantum information: {G}aussian states and
  beyond.
\newblock {\em Open Syst. Inf. Dyn.}, 21(01n02):1440001, 2014.

\bibitem{Wald}
Robert~M. Wald.
\newblock {\em General relativity}.
\newblock University of Chicago Press, Chicago, 1984.

\bibitem{Schr}
Erwin Schr\"odinger.
\newblock {\em Expanding universe}.
\newblock Cambridge University Press, Cambridge, 2011.

\bibitem{Felice_bini}
Fernando de~Felice and Donato Bini.
\newblock {\em Classical measurements in curved space-times}.
\newblock Cambridge University Press, Cambridge, 2010.

\bibitem{Chandra1983}
Subrahmanyan Chandrasekhar.
\newblock {\em The mathematical theory of black holes}.
\newblock Oxford University Press, Oxford, 1983.

\bibitem{Cramer}
Harald Cram\'er.
\newblock {\em Mathematical methods of statistics}.
\newblock Princeton University Press, Princeton, 1999.

\bibitem{squeezing}
Henning~Vahlbruch \textit{et al.}
\newblock Observation of squeezed light with 10-db quantum-noise reduction.
\newblock {\em Phys. Rev. Lett.}, 100:033602, 2008.

\bibitem{IAG1999}
{IERS} {N}umerical {S}tandards, {IAG1999}.
\newblock \url{http://hpiers.obspm.fr/eop-pc/models/constants.html}.

\bibitem{newexp}
K.U. Schreiber, T.~Kl\"ugel, J.-P.R. Wells, R.B. Hurst, and A.~Gebauer.
\newblock How to detect the {C}handler and the annual wobble of the {E}arth
  with a large ring laser gyroscope.
\newblock {\em Phys. Rev. Lett.}, 107:173904, 2011.

\bibitem{oldexp}
R.~Anderson, H.R. Bilger, and G.E. Stedman.
\newblock Sagnac effect: a century of {E}arth-rotated interferometers.
\newblock {\em Am. J. Phys.}, 62:975, 1994.

\bibitem{Bonato:Tomaello:2009}
C.~Bonato, A.~Tomaello, V.~Da Deppo, G.~Naletto, and P.~Villoresi.
\newblock Feasibility of satellite quantum key distribution.
\newblock {\em New Journal of Physics}, 11(4):045017, 2009.

\bibitem{Vallone:Bacco:2015}
Giuseppe Vallone, Davide Bacco, Daniele Dequal, Simone Gaiarin, Vincenza
  Luceri, Giuseppe Bianco, and Paolo Villoresi.
\newblock Experimental satellite quantum communications.
\newblock {\em Phys. Rev. Lett.}, 115:040502, Jul 2015.

\bibitem{losses}
Sebastian~P. Kish and Timothy~C. Ralph.
\newblock Estimating spacetime parameters with a quantum probe in a lossy
  environment.
\newblock {\em Phys. Rev. D}, 93:105013, 2016.

\bibitem{opt_phot}
Timothy C.~Ralph Peter P.~Rohde and Michael~A. Nielsen.
\newblock Optimal photons for quantum-information processing.
\newblock {\em Phys. Rev. A}, 72:052332, 2005.

\bibitem{quantum_sats}
Elizabeth Gibney.
\newblock Chinese satellite is one giant step for the quantum internet.
\newblock {\em Nature}, 535:478--479, July 2016.

\end{thebibliography}

\end{document}